\documentclass[10pt,twocolumn,showpacs,longbibliography,amsmath,amssymb,prd,floatfix,superscriptaddress]{revtex4-2}

\usepackage{graphicx}
\usepackage{dcolumn}
\usepackage{bm}
\usepackage{color}
\usepackage{txfonts}
\usepackage{microtype}
\usepackage{epstopdf}
\newcommand\numberthis{\addtocounter{equation}{1}\tag{\theequation}}

\usepackage{epstopdf}

\begin{document}

%\preprint{AIP/123-QED}

\title{Angular modulation of nonlinear Breit-Wheeler yield by vacuum dichroism}

\author{Jia-Ding Chen}
\affiliation{Department of Physics, Shanghai Normal University, Shanghai 200234, China}
\author{Ya-Nan Dai}
\affiliation{Shenzhen Key Laboratory of Ultraintense Laser and Advanced Material Technology, Center for Intense Laser Application Technology, and College of Engineering Physics, Shenzhen Technology University, Shenzhen 518118, China}
\author{Kai-Hong Zhuang}
\affiliation{Department of Physics, Shanghai Normal University, Shanghai 200234, China}
\author{Jing-Jing Jiang}
\affiliation{Department of Physics, Shanghai Normal University, Shanghai 200234, China}
\author{Baifei Shen}
\affiliation{Department of Physics, Shanghai Normal University, Shanghai 200234, China}
\author{Yue-Yue Chen}
\email{yue-yue.chen@shnu.edu.cn}
\affiliation{Department of Physics, Shanghai Normal University, Shanghai 200234, China}
\date {\today}

\begin{abstract}
Vacuum polarization is numerically investigated for the interaction between a GeV electron beam and a counterpropagating ultraintense laser pulse in the quantum radiation dominated-regime (QRDR). We identify a signal of vacuum polarization in pair density using a straightforward one-stage setup, circumventing the challenge of preparations of highly polarized probe photons or precise measurements of photon polarization.
In our scheme, most electrons are scattered in the direction of laser propagation while emitting substantial linearly polarized gamma photons. These photons undergo vacuum birefringence and dichroism before decaying into electron-positron pairs via the nonlinear Breit-Wheeler process. We demonstrate that vacuum dichroism enhances the purity of linear polarization, which suppresses the overall yield of electron-positron pairs and allows energetic photons to penetrate deeper into the laser pulse. The pairs produced by these energetic photons are more likely to be deflected into small-angle regions rather than being reflected, leading to an enhancement of pair yield in forward scattering. The difference in positron yield may have potential applications in measuring vacuum polarization effect in future laser-particle experiments.

\end{abstract}

\maketitle

%\section{Introduction}
Vacuum birefringence, a long-predicted effect in quantum electrodynamics (QED), 
remains unobserved in terrestrial experiments utilising real photons \cite{zavattini2012measuring,ejlli2020pvlas,cadene2014vacuum,zavattini2006experimental}.  Several ongoing experiments aim to detect the ellipticity signal of vacuum birefringence using optical photons and long magnetic field \cite{zavattini2006experimental,zavattini2012measuring,ejlli2020pvlas,cadene2014vacuum,marklund2006nonlinear,di2012extremely}. Noteworthy examples include the PVLAS \cite{zavattini2012measuring,zavattini2006experimental} and BMV \cite{cadene2014vacuum} experiments, which, despite significant progress, have yet to overcome challenges related to background noise \cite{cadene2014vacuum,ejlli2020pvlas}. An alternative strategy involves using high-powered laser facilities to polarize the vacuum, with an XFEL serving as the probe (See for instance the planned   HIBEF \cite{schlenvoigt2016detecting} and SEL \cite{shen2018exploring,xu2020xfel} experiments). %Compared to vacuum magnetic birefringence, this setup is expected to yield a stronger signal due to the enhanced field strength and probe photons energy. 

%However, there is now evidence from polarisation measurements of strongly magnetised neutron stars that vacuum birefringence has been observed, and 

The nonlinearity of the QED vacuum becomes significantly more pronounced for high-energy photons and intense external field, sparking interest in detecting vacuum polarization (VP) with combination of high power laser facilities and gamma probe photons \cite{nakamiya2017probing,king2016vacuum,bragin2017high,macleod2023strong}. The use of a gamma-ray probe brings the interaction into the high-energy nonperturbative regime, where the Euler-Heisenberg effective Lagrangian method becomes invalid for describing vacuum birefringence \cite{bragin2017high}. In this case, the QED photon polarization operator in the presence of a strong background field must be employed %\cite{baier1998electromagnetic,baier1975interaction,becker1975vacuum,meuren2013polarization,torgrimsson2021loops,meuren2015polarization,meuren2015high,dinu2014vacuum,bragin2017high,torgrimsson2021loops}. 
%The QED polarization operator within the one-loop approximation has been extensively studied in Refs. 
\cite{baier1998electromagnetic,baier1975interaction,becker1975vacuum,meuren2013polarization,torgrimsson2021loops,bragin2017high,king2023strong}.
With the one-loop QED probability, a simulation method applicable to investigating high-energy vacuum polarization has been developed recently \cite{dai2024fermionic}. In the high-energy regime, the vacuum exhibit both birefringence and dichroism. The vacuum dichroism corresponds to the imaginary part of polarization operator, which is related to the pair production probability
via the optical theorem, and the vacuum birefringence corresponds to the real
part of the polarization operator. The imaginary part and real part is related via Kramers-Kronig relations \cite{hutchings1992kramers,toll1952dispersion}.
%[J. S. Toll, Ph.D. thesis, Princeton University, 1952.]. 
Therefore, the relationship between tree-level and loop processes implies that observing pair yield corresponds to detecting nonlinear vacuum birefringence \cite{adam2021measurement,borysov2022using}. 

Detecting VP with a gamma probe typically involves a two-stage setup: first, polarized gamma probes are generated via linear Compton scattering of a week laser \cite{nakamiya2017probing,bragin2017high,dai2024fermionic} or Bremsstrahlung in a diamond crystal radiator target \cite{borysov2022using,apyan2008coherent}; then,  the highly polarized gamma  photons are selected by angle to collide with a high-power laser. A change in photon polarization serves as evidence of VP. However, the first stage may introduce experimental errors and additional complexity to the experimental setup \cite{abramowicz2021conceptual}. %of measuring pair density $\sim 10\%$. 
Additionally, such experiments require gamma-ray polarimetry and face a challenge in separating photonic signal from background, such as laser photons and secondary emissions of produced pairs. 

To circumvent these challenges posed by photonic signals, %of measuring gamma photon polarization and background noise, 
one solution is to measure the polarized nonlinear Breit-Wheeler process \cite{borysov2022using,dai2024fermionic,adam2021measurement}. The vacuum birefringence affect the polarization of probe gamma photons, and consequently impact the spin polarization of produced pairs. The positrons (electrons) are easier to measure than photons within a photon background, and the experimental detection capacity for leptons spin polarization  (typically $\sim$ 0.5\% \cite{narayan2016precision}) is currently much higher than that for $\gamma$ polarization (typically $\lesssim$10\% \cite{ozaki2016demonstration}).  A measurement of pair polarization \cite{dai2024fermionic} or the pairs yield for different polarization states \cite{borysov2022using} may open a new way of testing vacuum birefringence.  %, avoiding the difficulties associated with photonic signals. 
{\color{black}Unfortunately, these proposals still rely on the two-stage setup and polarization measurements. 
The high-energy photons generated in the first-stage must be separated from the scattered electrons using a 1-meter-long dipole magnet before being directed to a strong-field interaction point (IP) located several meters downstream of the radiator. The angular spectrum of the generated photons follows $1/\gamma$, which corresponds to approximately $\sim 30\mu$rad for $\gamma = 16.5$ GeV$/m_e$. Consequently, the beam size at the IP is around 230 $\mu$m, significantly larger than the typical beam size of strong laser ($w_0 = 5 \sim 25\mu$m) \cite{abramowicz2021conceptual}. Therefore, only 0.3\% of the prepared photons fall within $\pm 25\mu$m at the IP, and this fraction decreases quadratically with decreasing laser spot size. Consequently, the pair production yield is quite low due to this spatial mismatch, which hinders accurate measurement of vacuum polarization. More importantly, the experimental feasibility of previous schemes remains restricted by the capacity for polarization measurement, highlighting the significant interest in identifying more viable physical observables.}
Recently, the STAR collaboration has reported that angular modulation of the \textit{linear} Breit–Wheeler pair creation yield in ultra-peripheral heavy-ion collision experiments could be attributed to vacuum birefringence, implying a possible detection of vacuum nonlinearity using unpolarized measurement \cite{adam2021measurement}.

%alter the polarization state of probe photons,
In this paper, we numerically investigate the collision between an ultra-relativistic unpolarized electron beam and a linearly polarized laser pulse. The electrons emit high-energy gamma photons via nonlinear Compton scattering, which subsequently annihilate into electron-positron pairs through the nonlinear Breit-Wheeler process. %We identified a signal of VP in the pair yield. 
Vacuum dichroism enhances the linear polarization of intermediate photons, resulting in a reduced pair production rate. In a reflection scenario, the angular distribution of the produced pairs is influenced by VP: the overall positron density decreases while the density concentrated in the small-angle region increases. Contrary to conventional approaches, where VP effects (VPE) are observed through changes in the polarization of photons or produced pairs, we examine the impact of QED vacuum nonlinearity on pair yield without requiring polarization measurement.

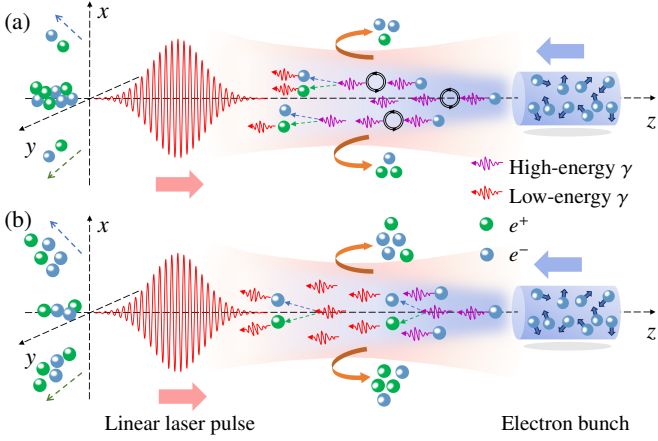
\begin{figure}
    \includegraphics[width=0.48\textwidth]{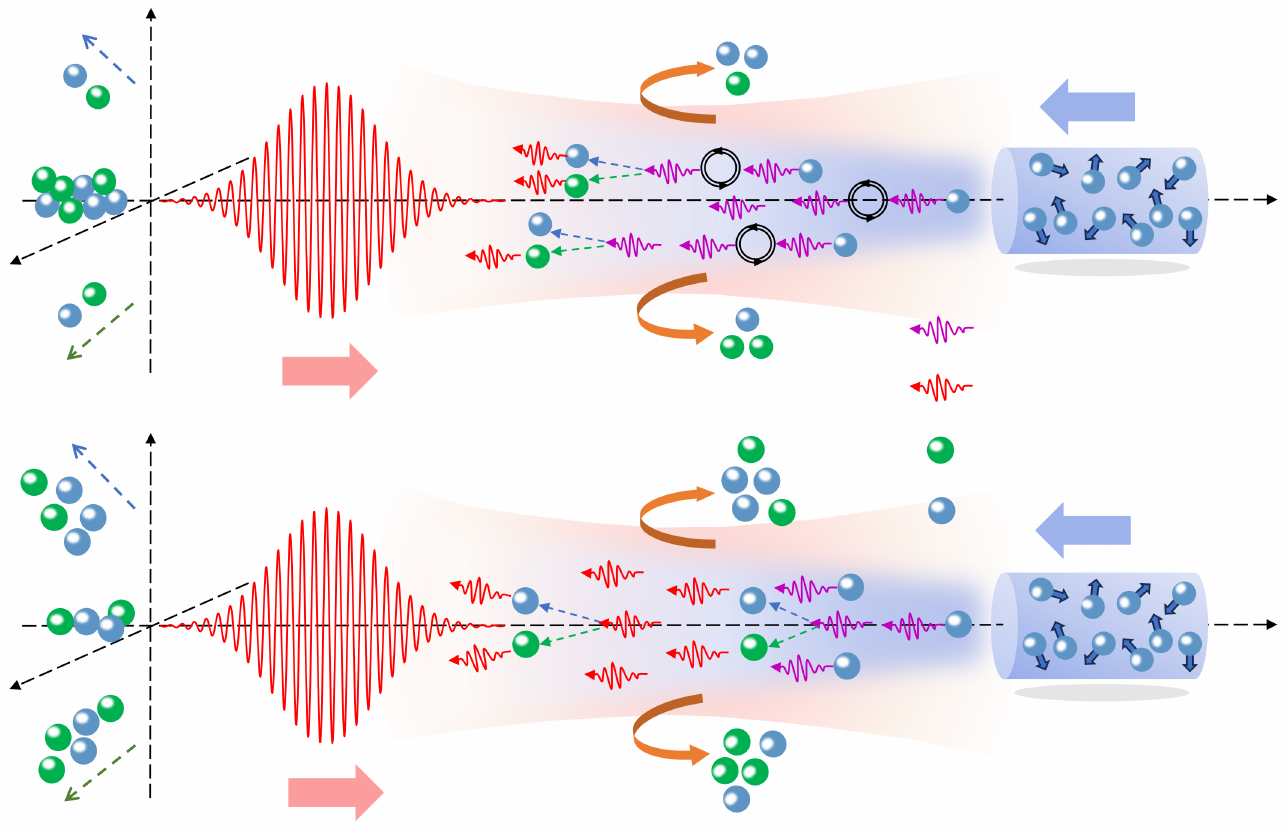}
       \begin{picture}(300,20)
    \put(-3,167){(a)}
    \put(-3,92){(b)}

    %\put(35,105){Linear laser pulse}
    \put(35,15){\footnotesize Linear laser pulse}

    %\put(185,155){Electron bunch}
    \put(185,15){\footnotesize Electron bunch}

    \put(32,171){$x$}
    \put(32,90){$x$}

    \put(5,40){$y$}
    \put(5,120){$y$}

    \put(240,130){$z$}
    \put(240,50){$z$}

    \put(188,112){\footnotesize High-energy $\gamma$}
    \put(188,101){\footnotesize Low-energy $\gamma$}
    \put(188,90){\footnotesize$e^{+}$}
    \put(188,79){\footnotesize$e^{-}$}
    \end{picture}
    \caption{ (a) Scheme for measuring vacuum polarization effect. The linearly polarized laser pulse head-on collision with an unpolarized GeV electron beam. {\color{black}The nonlinear Compton scattering and nonlinear Breit-Wheeler take place within the volume where the pulse
and the bunch overlaps. During the interaction,} most pairs are reflected in the $+z$ direction, while some continue to propagate forward. %The yield of pairs concentrated in the small-angle region is detected. 
(b) Same as (a), but with the vacuum polarization effect artificially turned off.  Without vacuum polarization, the yield of pairs in the small-angle region decreases.}
    \label{Fig.1d}
\end{figure}

%\section{THE THEORETICAL MODEL}
The high-energy VPE is explored through a recently developed Monte Carlo simulation method that incorporates both vacuum birefringence and vacuum dichroism \cite{dai2024fermionic}. At each time step,
the pair production is determined by the photon polarization-resolved pair production probability %derived  using QED operator method of Baier-Katkov under local-constant field approximation (LCFA), 
using the common algorithms \cite{dai2024fermionic,elkina2011qed,ridgers2014modelling,green2015simla,gonoskov2015extended,chen2019polarized,li2020production,dai2022photon,zhuang2023laser}.
% the photon Stokes parameters $\boldsymbol{\xi}=(\xi_1,\xi_2,\xi_3)$ are updated using instantaneous polarization basis vectors $\mathbf{e_1 = s - (n \cdot s)}$ and $\mathbf{e_2 = n \cdot s}$, where $\mathbf{s}$ represents the electron acceleration unit vector and $\mathbf{n}$ denotes the photon propagation direction \cite{li2020production,dai2022photon,zhuang2023laser,chen2022electron}.
%\cite{elkina2011qed,ridgers2014modelling,green2015simla,gonoskov2015extended,chen2019polarized,li2020production,dai2022photon,zhuang2023laser,chen2022electron}. The occurrence of pair production is determined using the photon-polarization-resolved probability via the standard stochastic algorithm \cite{zhuang2023laser}
%\cite{guo2020stochasticity,li2019ultrarelativistic,wan2020ultrarelativistic}
% The probability is derived  using QED operator method of Baier-Katkov under local-constant field approximation (LCFA) \cite{baier1998electromagnetic,chen2022electron,ritus1985quantum}. 
If pair production is rejected, the photon’s polarization state changes due to vacuum birefringence and vacuum dichroism. The vacuum birefringence is implemented via a polarization rotation between stokes parameters $\xi_1$ and $\xi_2$ \cite{dai2024fermionic,bragin2017high}:
%,dinu2014vacuum,torgrimsson2021loops,born2013principles,blum2012density
\begin{align}\label{vacuum birefringence}
\left(\begin{array}{c}
\xi_{1}^{f}\\
\xi_{2}^{f}
\end{array}\right)=\left(\begin{array}{cc}
\cos\varphi & \sin\varphi\\
-\sin\varphi & \cos\varphi
\end{array}\right)\left(\begin{array}{c}
\xi_{1}\\
\xi_{2}
\end{array}\right),
\end{align}
where $\varphi=\frac{\alpha m^{2}}{\omega^{2}}\Delta t\int d\varepsilon\frac{\textrm{Gi}'\left(\rho\right)}{\rho}$, with $\textrm{Gi}'\left(x\right)$ being the derivative of the Scorer function, $\rho=\left[\delta\left(1-\delta\right)\chi_{\gamma}\right]^{-2/3}$, $\delta=\varepsilon/\omega$, and $\chi_{\gamma}=\left|F_{\mu\nu}k^{\nu}\right|/mF_{cr}$ being the strong-field quantum parameter. Here, $\omega$ and $\varepsilon$  represent the energies of the photon and the produced electron, respectively. {\color{black}In this way, vacuum birefringence is incorporated at each time step, rather than being derived from the final yield of pairs via Hilbert transformation \cite{borysov2022using}. This approach is therefore more suitable for a time-dependent analysis of loop effects and investigating vacuum polarization when high-order processes (e.g. pair creation by secondary photons) could influence the final yield of pairs.}

Meanwhile, photons with different polarization states are absorbed at varying rates due to the photon polarization-dependent pair production probability. This selective effect on the initial photon polarization can cause polarization changes in the surviving particles, a phenomenon known as vacuum dichroism.  In our simulation, vacuum dichroism is modelled by updating photon polarization $\boldsymbol{\xi}$ to $\boldsymbol{\xi_f^{NP}}$ at each time step. $\boldsymbol{\xi_f^{NP}}$ represents the photon polarization state determined by the no-pair production probability \cite{dai2024fermionic}:
\begin{align*}\label{cain}
\boldsymbol{\xi_{f}^{NP}} & =\frac{\boldsymbol{\xi}\left(1-\underline{w}\Delta t\right)-\boldsymbol{\underline{f}}\Delta t}{1-\left\{ \underline{w}+\boldsymbol{\underline{f}}\cdot\boldsymbol{\xi}\right\} \Delta t},\\
\underline{w}&=\int\frac{\alpha m^{2}d\varepsilon}{\sqrt{3}\pi\omega^{2}}\left[\int_{z_{p}}^{\infty}dx\textrm{K}_{\frac{1}{3}}\left(x\right)+\frac{\varepsilon_{+}^{2}+\varepsilon^{2}}{\varepsilon\varepsilon_{+}}\textrm{K}_{\frac{2}{3}}\left(z_{p}\right)\right],\\
\boldsymbol{\underline{f}}&=-\int\frac{\alpha m^{2}d\varepsilon}{\sqrt{3}\pi\omega^{2}}\hat{e}_{3}\textrm{K}_{\frac{2}{3}}\left(z_{p}\right),\numberthis
\end{align*}
where $z_p=\frac{2}{3\chi_\gamma}\frac{\omega^2}{\varepsilon_+\varepsilon}$ with $\varepsilon_+$  being  positron energy, and $\hat{e}_{3}=(0,0,1)$.
It has been demonstrated that the no-pair production probability is equivalent to the $\alpha-$order loop probability related to vacuum dichroism \cite{dai2024fermionic,torgrimsson2021loops}. This algorithm for incorporating radiative corrections to photon self-energy aligns with the one used for electron self-energy \cite{li2023strong,li2022helicity}. 

Under the local-constant field approximation, the polarization operator is nonzero only when the external photon four-momentum is conserved \cite{meuren2013polarization}. Consequently, off-forward scattering does not occur in strong-field conditions, and the photon's momentum and energy remain unchanged at each time step.
%, leaving the photon’s angular distribution unaffected by vacuum polarization. However, as we discuss below, vacuum dichroism alters the angular distribution of the pair density.

%\section{Simulation RESULTS AND ANALYSIS}
We consider an ultra-relativistic electron beam with energy $\varepsilon_{0} = 10$ GeV colliding with a focused linearly polarized strong laser field with intensity $a_{0} = 350$ ($I\sim 10^{23}$ W/cm$^2$), as illustrated in Fig. 1. %Here, $E_0$ is the laser field amplitude and $\omega_0$ the frequency. The laser is polarized along the $x$ direction and propagates along the $z$ direction. 
The laser pulse has a duration of $\tau_{p} = 25$ fs, a wavelength of $\lambda_{0} = 800$ nm, and a focal radius of $w_{0} = 5\lambda_{0}$. The simulated electron beam consists of $10^{6}$ electrons uniformly distributed in the longitudinal direction. The length of the electron beam is $L = 5\lambda_{0}$, with a radius of $ \lambda_{0}$. The angular spread of the electron beam is $\Delta\theta = 2$ mrad, and the energy spread is $\Delta\varepsilon/\varepsilon = 5\%$, which are typical parameters for electron beams accelerated by laser wakefield acceleration and conventional accelerators.

\begin{figure}[t]
     \includegraphics[width=0.5\textwidth]{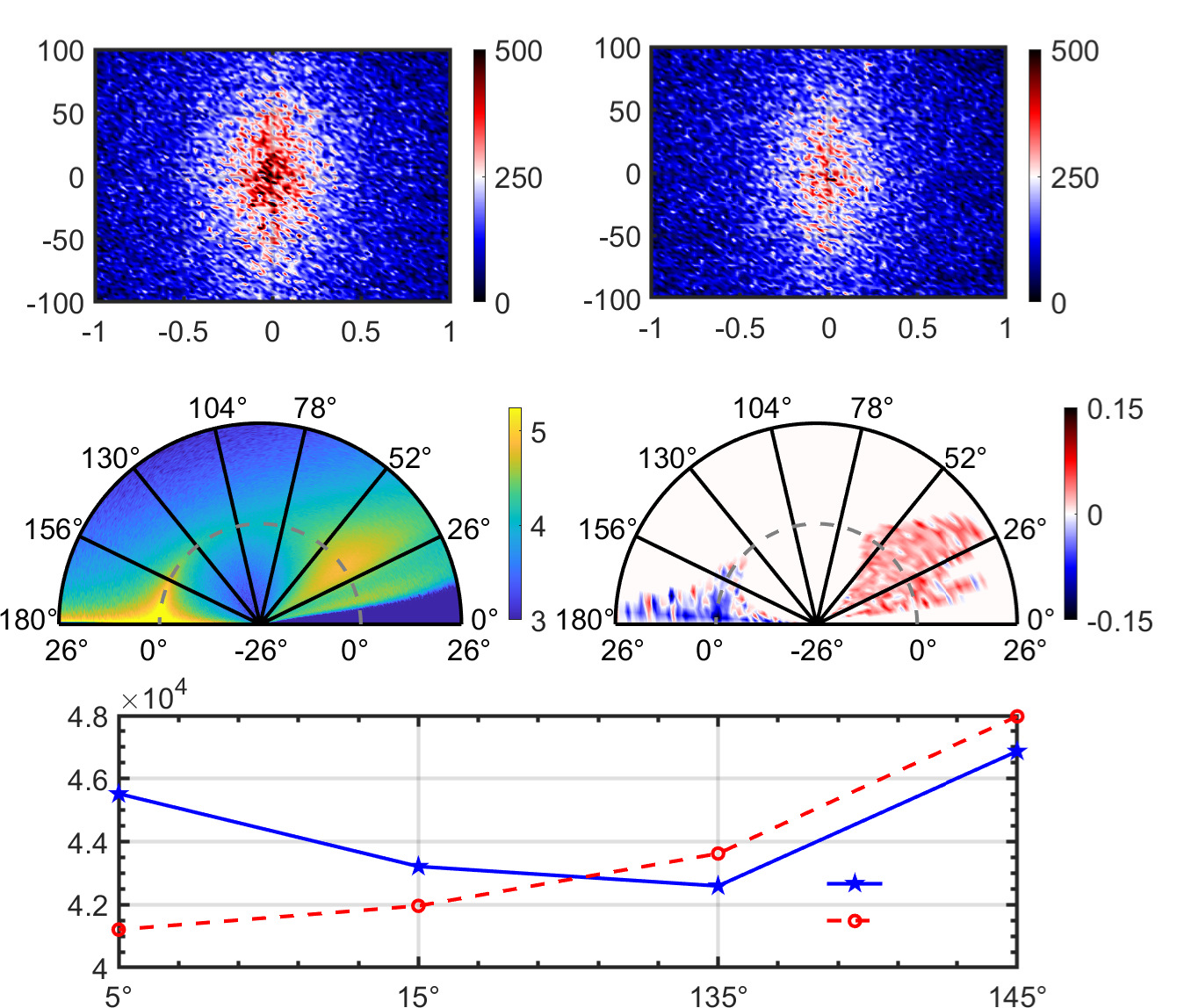}
       \begin{picture}(300,20)
    \put(25,217){\textcolor{white}{(a)}}
    \put(145,217){\textcolor{white}{(b)}}
    \put(15,150){(c)}
    \put(135,150){(d)}
    \put(30,72){(e)}
    
    %(a)
    \put(45,235){$\theta_{y}$(mrad)}
    \put(2,185){\rotatebox{90}{$\theta_{x}$(mrad)}}
    %(b)
    \put(165,235){$\theta_{y}$(mrad)}
    %\put(127,185){\rotatebox{90}{$\theta_{x}$(mrad)}}

    %c,d phi
    \put(53,88){$\phi$}
    \put(173,88){$\phi$}

    %e deflection angle particles
    \put(120,15){$\theta$}
    \put(5,53){\rotatebox{90}{$N_+$}}

    \put(190,45){\scriptsize VP}
    \put(190,37){\scriptsize NVP}
    
    \end{picture}
    \caption{The angular distribution of positron density $d^{2}N_{+}/d\theta_{x}d\theta_{y}$ within $\left|\theta_{x}\right|\lesssim100$ mrad and $\left|\theta_{y}\right|\lesssim1$ mrad, versus $\theta_x=\tan^{-1} p_x/p_z$ (mrad) and $\theta_y=\tan^{-1} p_y/p_z$ (mrad) for the cases: with  (a) and without (b) VPE. Angular distribution of positrons density $\text{log}_{10}(dN_+/d\Omega)$ (c) and density difference $R$ (d) vs the polar angle $\theta$ (degree, black solid scale) and azimuthal angle $\phi$ (degree, gray dashed scale). (e) The scaling law of positron yield $N_+$  within $\delta \theta=10^\circ$ and $\phi\in(-1^\circ,1^\circ)$, versus deflection angle $\Delta\theta=\pi-\theta$ for the cases: with (blue solid line) and without (red dashed line) VPE.  }
        \label{Fig.agl}
\end{figure}

\begin{figure}[b]
    \includegraphics[width=0.5\textwidth]{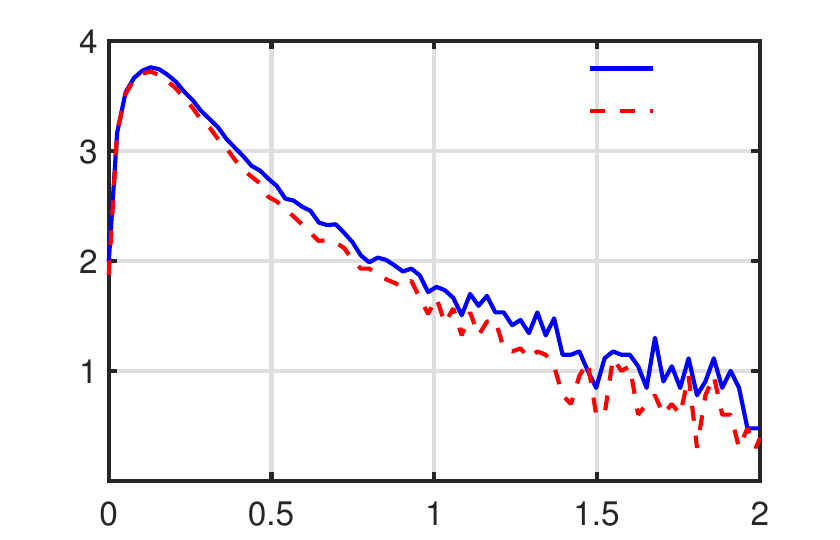}
    \begin{picture}(300,20)
   
    %（e） \footnotesize
    \put(120,10){ $\varepsilon_+$(GeV)}
    \put(10,80){\rotatebox{90}{ $\text{log}_{10}(dN_+/d\varepsilon_+)$}}
    
    \put(200,162){ VP}
    \put(200,149){NVP}
    
    \end{picture}
    \caption{The spectrum of positrons within $\left|\theta_{x}\right|\lesssim100$ mrad and $\left|\theta_{y}\right|\lesssim1$ mrad, with (blue solid line) and without VPE (red dashed line).}
       \label{Fig.spectrum}
\end{figure}
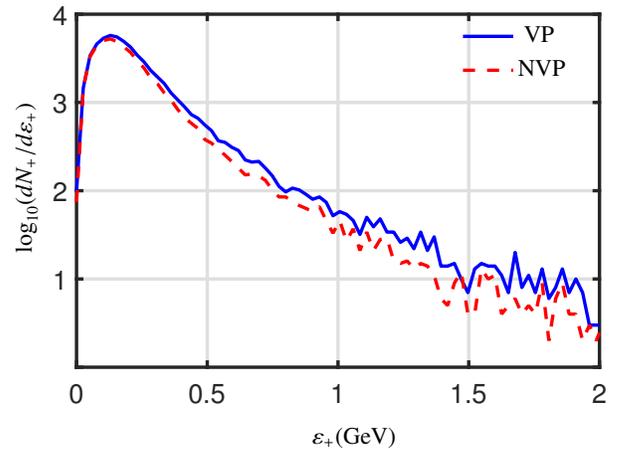

The simulation results for the angular distribution of positron density are shown in Fig. \ref{Fig.agl}.  The VP causes a decrease of positron yield from $N_\text{nvp}\approx1.2\times10^7$ to $N_\text{vp}\approx1.17\times10^7$, implying a relative difference of $R=\left(N_\text{nvp}-N_\text{vp}\right)/N_\text{vp}\approx 1.87\%$ induced by neglecting VPE. 
In the quantum radiation dominated-regime (QRDR), where $R_c=\alpha a_0\approx 1, \chi_e \gtrsim 1$ \cite{di2012extremely},  most of the produced pairs experience significant radiation and are reflected in the direction of laser propagation  ($\sim 70\%$), while a small portion of positrons  ($\sim 30\%$)  move forward into a small deflection angle, see Fig. \ref{Fig.agl} (c). VP reshapes the angular distribution of positrons, leading to more passing through the laser and fewer being reflected compared to when it is absent, see Fig. \ref{Fig.agl} (d) and (e). More intuitively, for small-angle positrons with $\left|\theta_{x}\right|\lesssim100$ mrad and $\left|\theta_{y}\right|\lesssim1$ mrad, the positron yield increases from $N_\text{nvp} \approx 5 \times 10^4$ to $N_\text{vp} \approx 5.8 \times 10^4$ in the case of VP, see Fig. \ref{Fig.agl} (a) and (b). The relative difference induced by VP is $R \approx 13.8\%$, which is significantly higher compared to $R \approx 1.87\%$ for the entire beam and $R \approx 2.3\%$ for the reflected positrons, see Fig. \ref{Fig.agl} (d). 
%Forward-scattering positrons are easier to select than reflected ones and carry stronger signals, therefore more subtitle for experimental measurement.

Moreover, besides its impact on positron yield, VP also affects the positron spectrum, as shown in Fig. \ref{Fig.spectrum}. VP results in the enhancement of positron yield in the high-energy regime ($\varepsilon_+\gtrsim 129$ MeV).  For the positrons with $\varepsilon_+ \gtrsim 1$ GeV, the average energy increases from $\overline{\varepsilon}_+^\text{nvp}=219$ MeV to $\overline{\varepsilon}_+^\text{vp}=237$ MeV, and the difference in pair yield increase to $R = 40\%$.
The differences in angular distribution and spectrum indicate that incorporating VP is crucial for accurately describing the pair properties. More importantly, the VP alters scaling law of the pair yield with respect to both the deflection angle [Fig. \ref{Fig.agl}(e)] and positron energy (Fig. \ref{Fig.spectrum}), offering a potential method for measuring VP. %using a conventional electron (positron) spectrometer \cite{poder2018experimental}.

%The spectral differences can act as an additional indicator of vacuum dichroism and can be detected using a conventional electron (positron) spectrometer \cite{poder2018experimental}.
 
\begin{figure}[]
 \hspace{-0.5cm} 
    \includegraphics[width=0.5\textwidth]{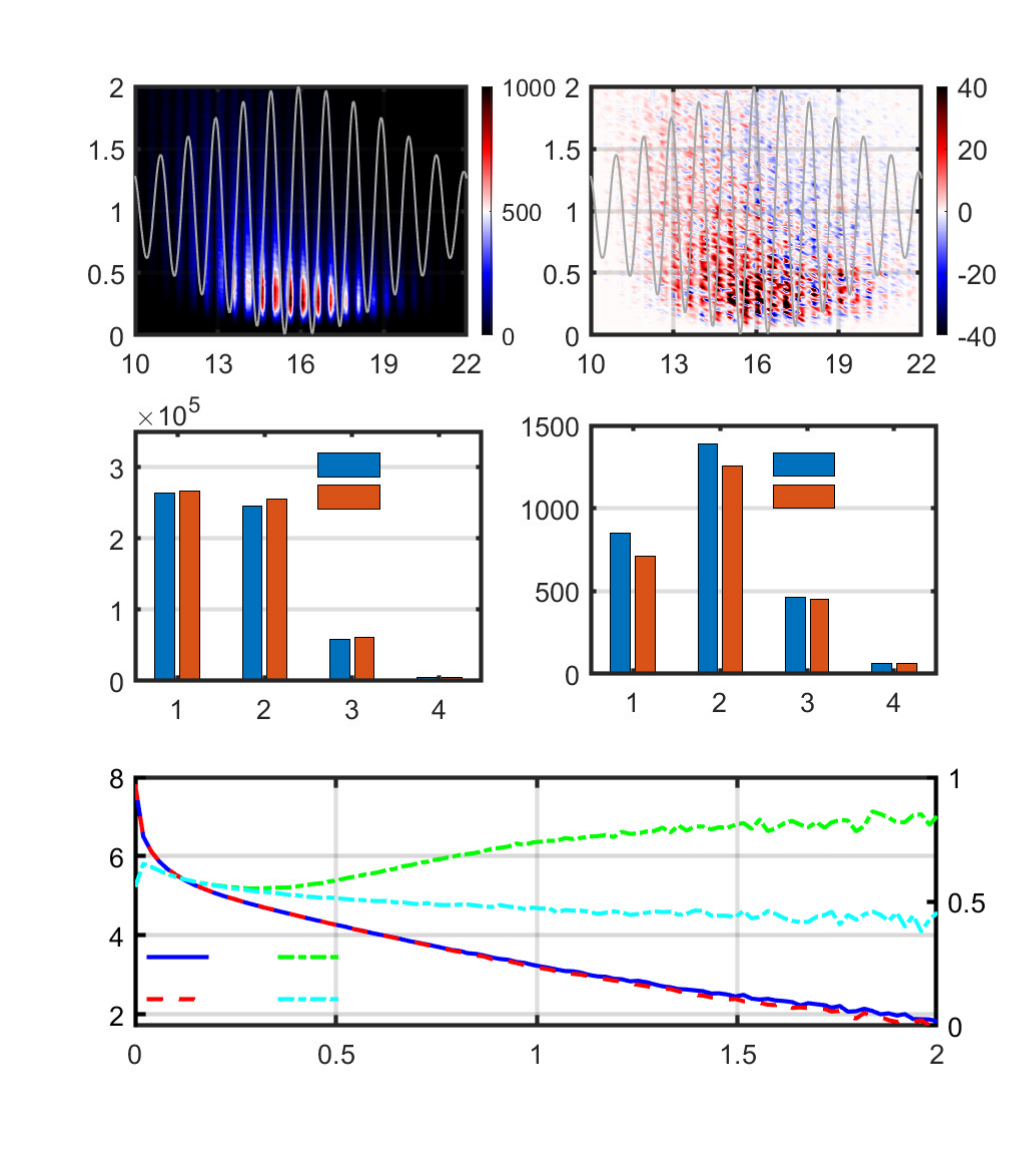}
\begin{picture}(300,20)
     \put(27,271){\textcolor{white}{(a)}}
    \put(140,271){(b)}
    \put(27,187){(c)}
    \put(140,187){(d)}
    \put(27,100){(e)}

    %（a）
    \put(55,205){$\varphi/(2\pi)$}
    \put(0,235){\rotatebox{90}{ $\omega$(GeV)}}

   % \put(5,230){\rotatebox{90}{ $\omega$(GeV)}}

    %（b）
    \put(167,205){$\varphi/(2\pi)$}
    %\put(117,230){\rotatebox{90}{$\omega$(GeV)}}

    %（c） %small
    \put(60,120){$n$}
    \put(0,160){\rotatebox{90}{$N_\gamma$}}
    \put(85,188){\scriptsize VP}
    \put(85,180){\scriptsize NVP}

    %（d）
    \put(170,120){ $n$}
    %\put(120,165){\rotatebox{90}{ N}}
    \put(198,188){\scriptsize VP}
    \put(198,180){\scriptsize NVP}

    %（e） \footnotesize
    \put(110,30){ $\omega$(GeV)}
    \put(0,50){\rotatebox{90}{ $\text{log}_{10}(dN_\gamma/d\omega)$}}
    \put(235,85){\rotatebox{-90}{ $\overline{\xi}_{3}$}}
    
    \put(42,67){\scriptsize VP}
    \put(75,67){\scriptsize VP}
    
    \put(42,57){\scriptsize NVP}
    \put(75,57){\scriptsize NVP}
    
    \end{picture}

    %\vspace{-30pt}
    \caption{The photon density $d^2N_\gamma/d\omega d\varphi$ with VPE (a) and the difference of pair yield $N_\text{nvp}-N_\text{vp}$ (b) versus pair production phase $\varphi(2\pi)$  and photon energy $\omega$ (GeV). (c) The total photon number $N_\gamma$ that decay into pairs versus photon generations $n$. (d) Same with (c) but for the photons that produce pairs within $\left|\theta_{x}\right|\lesssim100$ mrad and $\left|\theta_{y}\right|\lesssim1$ mrad.  (e) Photon spectra with (solid blue line) and without (red dashed line) VPE. The average photon polarization $\xi_3$ versus photon energy $\omega$ (GeV) with (green dotted line) and without (cyan dot-dashed line) VPE. }
       \label{Fig.tra}
\end{figure}
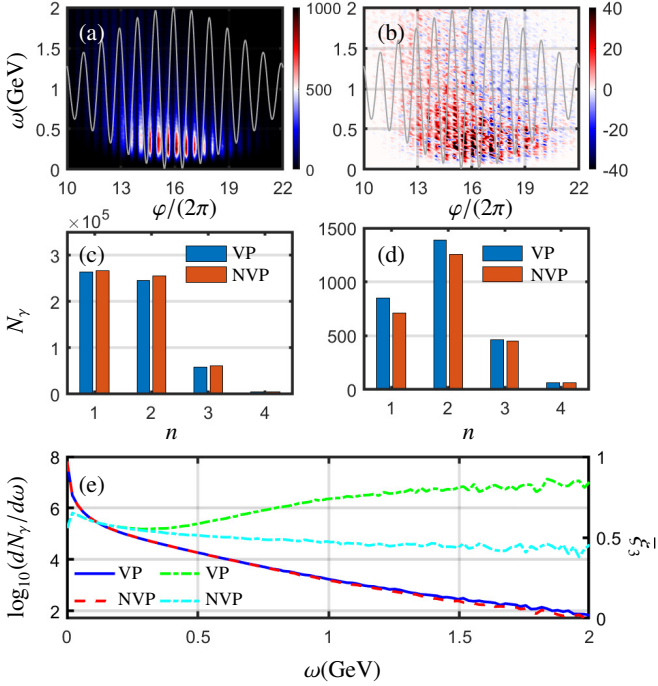

To analyse how VP influences the pair yield, we traced the trajectories of gamma photons that decay into pairs, see Fig. \ref{Fig.tra}. The electrons emit substantial photons at the laser front, which are primarily composed of photons with energies in the hundreds of MeV, see Fig.  \ref{Fig.tra} (a) and  (e) . 
The average polarization of the emitted photons by unpolarized electrons (positrons) can be estimated using \cite{dai2022photon,dai2024fermionic}
\begin{align}\label{xi3}
\xi_1&=\xi_2=0,\xi_{3}=\textrm{K}_{\frac{2}{3}}\left(z_{q}\right)\left[\frac{\varepsilon^{2}+\varepsilon'^{2}}{\varepsilon'\varepsilon}\textrm{K}_{\frac{2}{3}}\left(z_{q}\right)-\int_{z_{q}}^{\infty}dx\textrm{K}_{\frac{1}{3}}\left(x\right)\right]^{-1},
\
%\xi_{3}&=\frac{\textrm{K}_{\frac{2}{3}}\left(z_{q}\right)}{\frac{\varepsilon^{2}+\varepsilon'^{2}}{\varepsilon'\varepsilon}\textrm{K}_{\frac{2}{3}}\left(z_{q}\right)-\int_{z_{q}}^{\infty}dx\textrm{K}_{\frac{1}{3}}\left(x\right)},
\end{align}
where $z_q=\frac{2}{3}\frac{\omega}{\chi_e\varepsilon'}$ with $\varepsilon$ and $\varepsilon'$ being the electron (positron) energy before and after emission, respectively. 
For the majority of the emission, where the photon energy satisfies $\omega \ll \varepsilon$,  it follows that $\xi_1 = \xi_2 = 0$ and $\xi_3 \approx 0.5$, as indicated by Eq. (\ref{xi3}) [Fig. \ref{Fig.tra} (e)]. %These gamma photons are likely to decay into pairs in front of the laser, which are subsequently reflected due to radiation reaction and ponderomotive force.
%Generally speaking, the vacuum polarization suppresses the pair yield.
According to Eq. (\ref{vacuum birefringence}), vacuum birefringence has negligible effects on photons with $\xi_1 = \xi_2 = 0$. It is vacuum dichroism that plays a crucial role in altering the polarization of emitted photons as they travel through the laser. vacuum dichroism induces a change in $\xi_3$ following \cite{dai2024fermionic,bragin2017high}
\begin{align}
\frac{d\xi_{3}}{dt} & =\int\frac{\alpha m^{2}d\varepsilon}{\sqrt{3}\pi\omega^{2}}\left(1-\xi_{3}^{2}\right)\textrm{K}_{\frac{2}{3}}\left(z_{p}\right).
\end{align}
Photons initially generated with $\xi_3 \approx 0.5$ undergo an increased degree of polarization due to the influence of vacuum dichroism. Since the rate of variation is primarily governed by the photon energy through the $\textrm{K}_{\frac{2}{3}}\left(z_{p}\right)$ term, higher-energy photons acquire larger $\xi_3$ travelling through the laser field [Fig. \ref{Fig.tra} (e)].

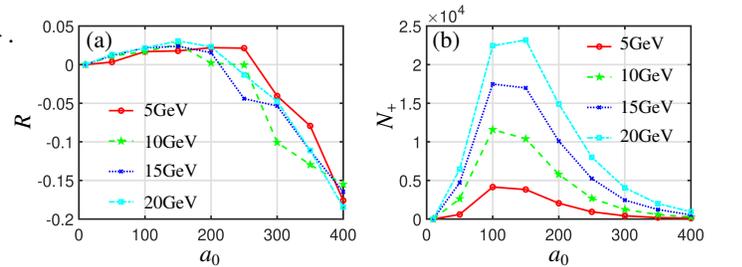
\begin{figure}[b]
     \includegraphics[width=0.5\textwidth]{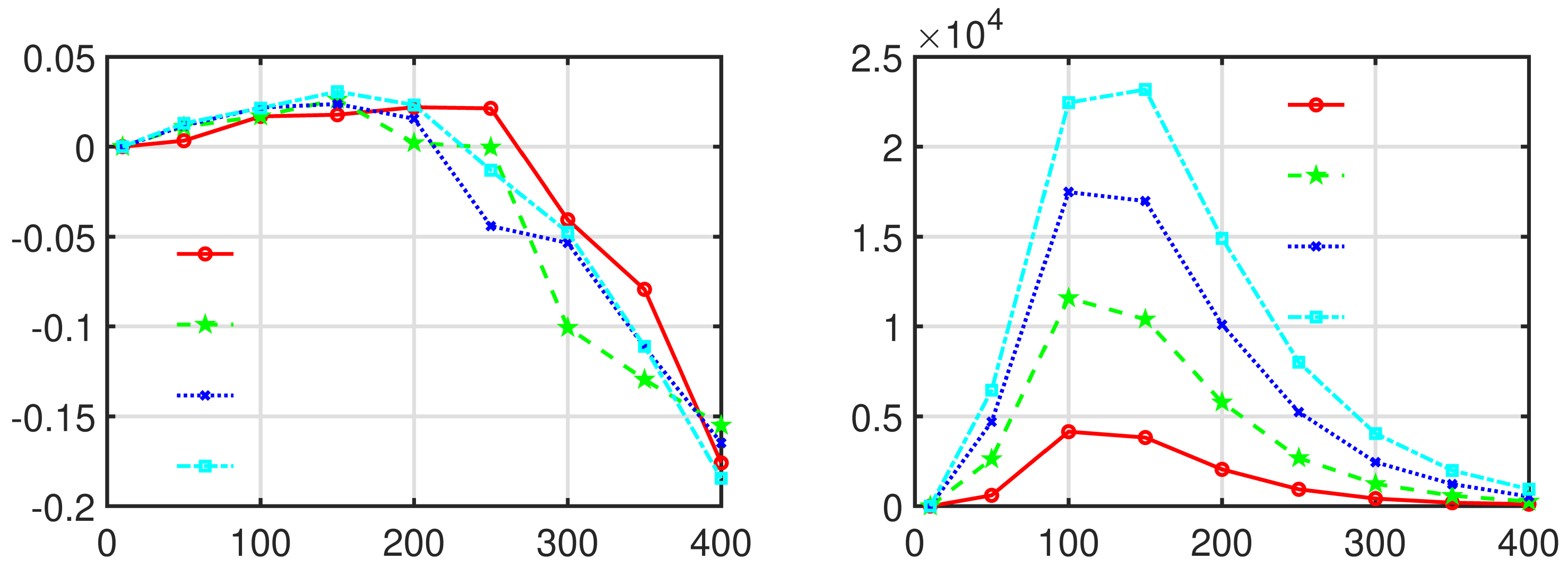}
     \begin{picture}(300,20)
    \put(20,96){(a)}
    \put(151,96){(b)}
    
    \put(63,15){$a_{0}$}
    \put(-7,65){\rotatebox{90}{$R$}}
 %a
    \put(42,70){\scriptsize 5GeV}
    \put(42,58){\scriptsize 10GeV}
    \put(42,47){\scriptsize 15GeV}
    \put(42,35){\scriptsize 20GeV}
%b
    \put(195,15){$a_{0}$}
    \put(130,65){\rotatebox{90}{$N_+$}}
    \put(222,95){\scriptsize 5GeV}
    \put(222,83){\scriptsize 10GeV}
    \put(222,71){\scriptsize 15GeV}
    \put(222,60){\scriptsize 20GeV}
    
    \end{picture}
    \caption{The scaling law of $R$ (a) and pair number $N_+$ (b) versus $a_0$ for electrons with initial energy of 5GeV (red solid line), 10GeV (green dashed line), 15GeV (blue dotted line) and 20 GeV (cyan dot-dashed line), for the pairs with $\left|\theta_{x}\right|\lesssim100$ mrad and $\left|\theta_{y}\right|\lesssim1$ mrad.}
        \label{Fig.scale}
\end{figure}

Furthermore, the probability of pair production depends on photon polarization \cite{chen2022electron}:
\begin{align}
dW^P & =\frac{\alpha m^{2}d\varepsilon}{\sqrt{3}\pi\omega^{2}}\left\{ \int_{z_{p}}^{\infty}dx\textrm{K}_{\frac{1}{3}}\left(x\right)+\frac{\varepsilon_{+}^{2}+\varepsilon^{2}}{\varepsilon\varepsilon_{+}}\textrm{K}_{\frac{2}{3}}\left(z_{p}\right)-\xi_{3}\textrm{K}_{\frac{2}{3}}\left(z_{p}\right)\right\}.
\end{align}
Consequently, high-energy photons can penetrate deeper into the laser to produce pairs at the tail, due to the  enhancement of $\xi_3$. As shown in Fig. \ref{Fig.tra} (b), incorporating VPE induces a phase delay in pair production, reducing the positron yield at the laser front while enhancing it at the laser tail. The positrons generated in the front are likely to be reflect due to the dramatic radiation reaction near laser peak. In contrast, these  generated in the pulse tail have higher energy and experience less radiation reaction. As a result, they tend to move forward with a small deflection angle. Consequently, both the positron yield and energy within $\left|\theta_{x}\right|\lesssim100$ mrad and $\left|\theta_{y}\right|\lesssim1$ mrad increase when VPE is considered, see Figs. \ref{Fig.agl} and \ref{Fig.spectrum}. The phase delay induced by vacuum dichroism is evident in the reflection geometry. Without it, pairs produced at different phases would mix within a small deflection angle, diminishing the strength of the VP signal.

% reflection geometry can isolate pairs whose parent photons undergo strong vacuum polarization, thereby enhancing the signal of vacuum polarization.
%In contrast, low-energy photons are less affected by vacuum polarization.  These pairs are created with lower energy and are more likely to be deflected with large angle. 
%The reflection geometry could single out pairs generated by energetic photons, which undergoes stronger  thereby enhancing the signal of vacuum polarization.

Meanwhile, even though multiple emissions occur in the QRDR, the difference in pair yield is primarily driven by the first and secondary photon emissions, see Fig. \ref{Fig.tra} (c) and (d). Compared with the third and forth generations, the first two generations are more energetic and are produced in an earlier phase, resulting in a larger $\xi_3$ and consequently a more pronounced difference in pair yield. %For the fist two generations, it is evident  that while the total pair yield decreases with the inclusion of VP [Fig. \ref{Fig.tra} (c)], there is a more pronounced increase in yield at small angles [Fig. \ref{Fig.tra} (d)].  %The relative difference in pair yield is particularly pronounced for particles at these small angles.

%\section{Impacts of the laser and electron-beam parameters on the signal of vacuum polarization}

We further studied the impacts of the laser and electron beam parameters on the relative difference $R$, as shown in Fig. \ref{Fig.scale} (a). The relative difference in pair yield depends on the laser intensity, while there is no particularly strong correlation with respect to the energy of the initial electrons. For $a_0 \lesssim 100$, we have $0<R<3\%$, indicating that the pair yield is larger when VP is ignored and that the relative difference is rather small. This is because the particles in a relatively week field tend to move forward rather than being reflected. Without reflection geometry, %pairs produced at different phases are mixed together, and the phase delay induced by VP has a negligible effect. 
the copious pairs produced at the laser front, where $R > 0$, are mixed with those from the laser tail, where $R < 0$, results in a small positive $R$. As the laser intensity increases to $100 < a_0 < 200$, we enter a relatively stable region where $R \approx 3\%$ and the pair yield reaches its peak at $N \approx 2.3 \times 10^4$, implying that VP measurement may be feasible with current laser technology. 
As $a_0$ further increases to the QRDR, pairs produced at the laser front lose substantial energy and are reflected, isolating the pairs generated by energetic photons at the laser tail. These pairs exhibit a stronger negative $R$.
%Consequently, the signal of VP $R$ enhances as $a_0$ increases. 
It is important to note that, although $R$ is significantly enhanced in an extremely intense laser field, the pair yield at small angles decreases [Fig. \ref{Fig.scale} (b)]. This results in larger statistical errors and poses challenges for experimental measurement. %Thus, while measuring VP may be feasible with current laser technology, a stronger signal could be achieved with more intense lasers, though this may come at the cost of reduced pair density.

\begin{figure}[]
    \includegraphics[width=0.5\textwidth]{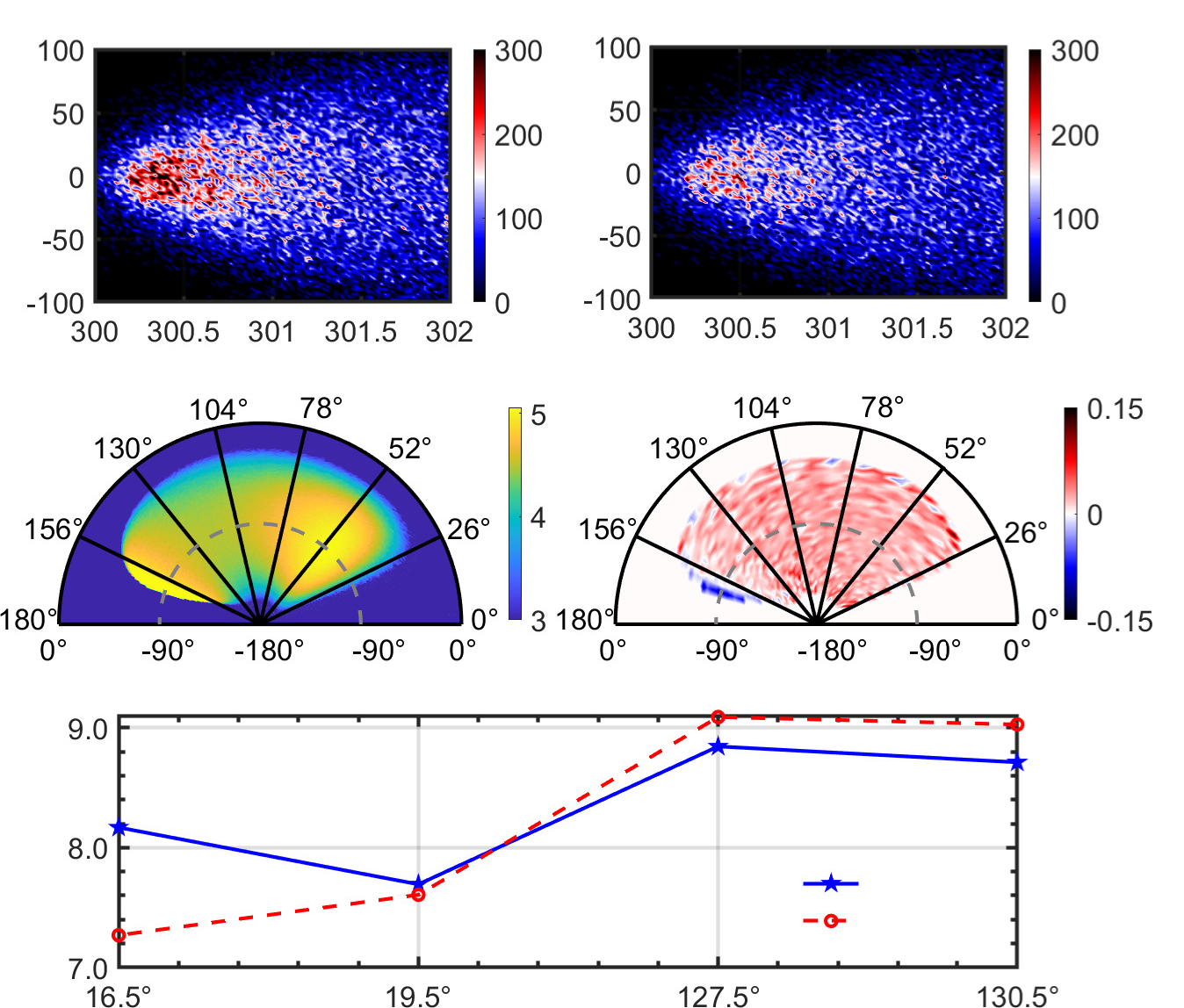}
    \begin{picture}(300,20)
    \put(25,217){\textcolor{white}{(a)}}
    \put(145,217){\textcolor{white}{(b)}}
    \put(15,150){(c)}
    \put(135,150){(d)}
    \put(30,72){(e)}
    
    %(a)
    \put(45,235){$\theta_{y}$(mrad)}
    \put(2,185){\rotatebox{90}{$\theta_{x}$(mrad)}}
    %(b)
    \put(165,235){$\theta_{y}$(mrad)}
    %\put(127,185){\rotatebox{90}{$\theta_{x}$(mrad)}}

    %c,d phi
    \put(53,88){$\phi$}
    \put(173,88){$\phi$}

    %e deflection angle particles
    \put(120,15){$\theta$}
    \put(5,53){\rotatebox{90}{$N_+$}}
    \put(25,85){\scriptsize $\times10^{3}$}

    \put(190,45){\scriptsize VP}
    \put(190,37){\scriptsize NVP}
    
    \end{picture}
    \caption{{\color{black}The angular distribution of positron density $d^{2}N_{+}/d\theta_{x}d\theta_{y}$ within $\left|\theta_{x}\right|\lesssim100$ mrad and $300\lesssim\theta_{y}\lesssim302$ mrad, versus $\theta_x=\tan^{-1} p_x/p_z$ (mrad) and $\theta_y=\tan^{-1} p_y/p_z$ (mrad) for the cases: with  (a) and without (b) VPE. Angular distribution of positrons density $\text{log}_{10}(dN_+/d\Omega)$ (c) and density difference $R$ (d) vs the polar angle $\theta$ (degree, black solid scale) and azimuthal angle $\phi$ (degree, gray dashed scale). (e) The scaling law of positron yield $N_+$  within $\delta \theta=3^\circ$ and $\phi\in(-91^\circ,-89^\circ)$, versus deflection angle $\Delta\theta=\pi-\theta$ for the cases: with (blue solid line) and without (red dashed line) VPE.  The collision angle between the laser pulse and the initial electron beam is $\theta_c=\tan^{-1}p_y/p_z=17.2^\circ$.}}
        \label{Fig.tilt}
\end{figure}

{\color{black}Furthermore, we conducted a simulation with the collision angle $\theta_c=17.2^\circ$,  as employed in the LUXE experiment.
The simulation results for the angular distribution of positron density are shown in Fig. \ref{Fig.tilt}. Although the total pair yield decreases slightly from $N_\text{vp}\approx1.17\times10^7$ to $N_\text{vp}\approx9.0\times10^6$  as the collision angle increases from $\theta_c=0^\circ$ to $\theta_c=17.2^\circ$, the relative difference $R$ induced by vacuum polarization is barely changed. Specifically, for the oblique collision at $\theta_c=17.2^\circ$, the VP causes a decrease of positron yield from $N_\text{nvp}\approx9.2\times10^6$ to $N_\text{vp}\approx9.0\times10^6$, implying a relative difference of $R=\left(N_\text{nvp}-N_\text{vp}\right)/N_\text{vp}\approx 1.94\%$ induced by neglecting VPE. This observable $R=1.94\%$ at $\theta_c=17.2^\circ$  is roughly the same with $R=1.87\%$ for $\theta_c=0^\circ$. Moreover, in head-on collisions, the pairs are concentrated at $\theta_y=0$, whereas they shift toward $\theta_y=300$mrad in oblique collisions, see Fig. \ref{Fig.tilt} (a)-(c). However, in both cases, VP reshapes the angular distribution of positrons, leading to more passing through the laser and fewer being reflected compared to when it is absent, see Fig. \ref{Fig.tilt} (d) and (e). More intuitively, for small-angle positrons with $\left|\theta_{x}\right|\lesssim100$ mrad and $300\lesssim\theta_{y}\lesssim302$ mrad, the positron yield increases from $N_\text{nvp} \approx 2.1 \times 10^4$ to $N_\text{vp} \approx 2.5 \times 10^4$ in the case of VP, see Fig. \ref{Fig.tilt} (a) and (b). The relative difference induced by VP is $R \approx 15.7\%$, which is roughly consistent with $R \approx 13.8\%$ in the head-on collision scenario.  Therefore, our scheme is robust against variation in collision angle. {\color{black}To ensure optimal temporal and spatial alignment during the interaction, the laser jitter should be kept smaller than the Rayleigh length. In a recent collision experiment, the temporal jitter was approximately 11 fs \cite{mirzaie2024all}, which is significantly smaller than the Rayleigh length $z_R=\pi w_0^2/\lambda_0=210$ fs for $w_0=5\lambda_0$. } }

%\section{Conclusion}
In conclusion, we explored VPE through nonlinear Compton scattering of an unpolarized electron beam with a linearly polarized strong laser. Unlike the traditional two-stage approach, which requires the preparation of a highly polarized and well-collimated photon beam, our method simplifies the process by utilizing a single collision with an unpolarized electron beam.
During the collision, the vacuum dichroism induces an increase in $\xi_3$ for the emitted gamma photons, resulting in a decrease in the overall pair yield ($\sim$1.87\%) and causing a phase delay in the pair production for high-energy photons. In a reflection scenario, this phase delay further alters the angular distribution of the produced pairs, particularly in the small-angle regions, with $\left|\theta_{x}\right|\lesssim100$ mrad and $\left|\theta_{y}\right|\lesssim1$ mrad. In this angle region, vacuum dichroism enhances both the yield and energy of the  produced pairs. The relative difference from neglecting VPE can reach up to $40\%$, highlighting  the importance of including VPE in high-field QED simulations. Additionally,  measuring the energy spectrum and angular distribution of pairs provides information about the intermediate photon polarization, revealing the effects of vacuum polarization without the challenges associated with polarization measurements. Thus, our approach, which relies on pair yield as a signal, may provide a more feasible method for measuring VP.

{\it Acknowledgement:}
We gratefully acknowledge helpful dis- cussions with Karen Z. Hatsagortsyan. This work is supported by the National Key R\&D Program of China (Grant No. 2021YFA1601700), and the National Natural Science Foundation of China (Grants No.12474312 and No. 12074262).

\bibliography{prp}

\end{document}